\documentclass[useAMS,usenatbib]{mn2e}
\usepackage{epsfig}

\newif\ifAMStwofonts

\newif\ifoldfss
\newif\ifnfssone
\newif\ifnfsstwo
\def\f@s@s{}

\def\hvect#1{{\hat{\mathbfit{#1}}}} 
\def\vect#1{{\mathbfit{#1}}} 

\def\et{{\it et~al.~}}
\def\deg{\ifmmode^\circ _\cdot\else$^\circ _ \cdot$\fi }    
\def\degg{\ifmmode^\circ \else$^\circ $\fi } 

\ifoldfss
  \ifCUPmtlplainloaded \else
    \NewTextAlphabet{textbfit} {cmbxti10} {}
    \NewTextAlphabet{textbfss} {cmssbx10} {}
    \NewMathAlphabet{mathbfit} {cmbxti10} {} 
    \NewMathAlphabet{mathbfss} {cmssbx10} {} 
  \fi
  \ifAMStwofonts
    \ifCUPmtlplainloaded \else
      \NewSymbolFont{upmath} {eurm10}
      \NewSymbolFont{AMSa} {msam10}
      \NewMathSymbol{\upi}     {0}{upmath}{19}
      \NewMathSymbol{\umu}     {0}{upmath}{16}
      \NewMathSymbol{\upartial}{0}{upmath}{40}
      \NewMathSymbol{\leqslant}{3}{AMSa}{36}
      \NewMathSymbol{\geqslant}{3}{AMSa}{3E}

      \let\leq=\leqslant \let\le=\leqslant
       
    \fi
  \fi
\fi 

\ifnfssone
  \newmathalphabet{\mathit}
  \addtoversion{normal}{\mathit}{cmr}{m}{it}
  \addtoversion{bold}{\mathit}{cmr}{bx}{it}
  \newmathalphabet{\mathbfit} 
  \addtoversion{normal}{\mathbfit}{cmr}{bx}{it}
  \addtoversion{bold}{\mathbfit}{cmr}{bx}{it}
  \newmathalphabet{\mathbfss} 
  \addtoversion{normal}{\mathbfss}{cmss}{bx}{n}
  \addtoversion{bold}{\mathbfss}{cmss}{bx}{n}
  \ifAMStwofonts
    \ifCUPmtlplainloaded \else
      \UseAMStwoboldmath
      \makeatletter
      \new@mathgroup\upmath@group
      \define@mathgroup\mv@normal\upmath@group{eur}{m}{n}
      \define@mathgroup\mv@bold\upmath@group{eur}{b}{n}
      \edef\UPM{\hexnumber\upmath@group}
      \new@mathgroup\amsa@group
      \define@mathgroup\mv@normal\amsa@group{msa}{m}{n}
      \define@mathgroup\mv@bold\amsa@group{msa}{m}{n}
      \edef\AMSa{\hexnumber\amsa@group}
      \makeatother
      \mathchardef\upi="0\UPM19
      \mathchardef\umu="0\UPM16
      \mathchardef\upartial="0\UPM40
      \mathchardef\leqslant="3\AMSa36
      \mathchardef\geqslant="3\AMSa3E

      \let\leq=\leqslant \let\le=\leqslant

    \fi
  \fi
\fi 

\ifnfsstwo
  \DeclareMathAlphabet{\mathbfit}{OT1}{cmr}{bx}{it}
  \SetMathAlphabet\mathbfit{bold}{OT1}{cmr}{bx}{it}
  \DeclareMathAlphabet{\mathbfss}{OT1}{cmss}{bx}{n}
  \SetMathAlphabet\mathbfss{bold}{OT1}{cmss}{bx}{n}
  \ifAMStwofonts
    \ifCUPmtlplainloaded \else
      \DeclareSymbolFont{UPM}{U}{eur}{m}{n}
      \SetSymbolFont{UPM}{bold}{U}{eur}{b}{n}
      \DeclareSymbolFont{AMSa}{U}{msa}{m}{n}
      \DeclareMathSymbol{\upi}{0}{UPM}{"19}
      \DeclareMathSymbol{\umu}{0}{UPM}{"16}
      \DeclareMathSymbol{\upartial}{0}{UPM}{"40}
      \DeclareMathSymbol{\leqslant}{3}{AMSa}{"36}
      \DeclareMathSymbol{\geqslant}{3}{AMSa}{"3E}

      \let\leq=\leqslant \let\le=\leqslant

    \fi
  \fi
\fi 

\ifCUPmtlplainloaded \else
  \ifAMStwofonts \else 
    \def\upi{\pi}
    \def\umu{\mu}
    \def\upartial{\partial}
  \fi
\fi

\title[All-sky component separation]
{All-sky component separation in the presence of
 anisotropic noise and dust temperature variations}

\author[V.~Stolyarov et al.]{V.~Stolyarov$^{1,2}$, M.P.~Hobson$^2$, 
A.N.~Lasenby$^2$ and R.B.Barreiro$^3$\\
$^1$ Institute of Astronomy, Madingley Road, Cambridge CB3 0HA, UK\\
$^2$ Astrophysics Group, Cavendish Laboratory, Madingley Road, 
Cambridge CB3 0HE, UK\\
$^3$ Instituto de Fisica de Cantabria,
Avda. de los Castros, s/n 39005, Santander, Spain}

\date{Accepted ---. Received \today; in original form \today}

\pagerange{\pageref{firstpage}--\pageref{lastpage}}
\pubyear{2002}

\begin{document}

\maketitle

\label{firstpage}

\begin{abstract}

We present an extension of the harmonic-space maximum-entropy
component separation method (MEM) for multi-frequency CMB observations
that allows one to perform the separation with more plausible
assumptions about the receiver noise and foreground astrophysical
components. Component separation is considered in the presence of
spatially-varying noise variance and spectral properties of the
foreground components. It is shown that, if not taken properly into
account, the presence of spatially-varying foreground spectra, in
particular, can severely reduce the accuracy of the component
separation. Nevertheless, by extending the basic method to accommodate
such behaviour and the presence of anisotropic noise, we find that the
accuracy of the component separation can be improved to a level
comparable with previous investigations in which these effects
were not present.

\end{abstract}


\section{Introduction}
\label{intro}

An important stage in the reduction of CMB anisotropy data is the
separation of the astrophysical and cosmological components. Several
techniques have been suggested, including blind (Baccigalupi \et 2000,
Maino \et 2002) and non-blind (Hobson \et 1998, Bouchet and Gispert
1999, Stolyarov \et 2002) approaches. Non-blind methods, such as the
maximum-entropy method (MEM) or Wiener filtering, allow one to use all
available prior information about the components in the separation
process.

A detailed description of the harmonic-space MEM approach for flat
patches of the sky was described by Hobson \et (1998), and was
extended later to the sphere by Stolyarov \et (2002; hereinafter S02).
Accounting for the presence of point sources was discussed by Hobson
\et (1999) for the flat patches and point source detection on the full
sky maps using Spherical Mexican Hat Wavelets (SMHW) was analyzed by
Vielva \et (2003). A joint technique using both SMHW and MEM for the
flat case was investigated by Vielva \et (2001), and for the spherical
case it will be described in a forthcoming paper by Stolyarov \et (in
preparation).  The method has also been used to construct simulated
all-sky catalogues of the thermal Sunyaev-Zel'dovich effect in galaxy
clusters (Geisb\"usch, Kneissl and Hobson, in preparation).

The separation tests described in previous articles were performed
making some simplifying assumptions. In particular, the receiver noise
was assumed uncorrelated and statistically homogeneous over the sky,
which is a reasonable approximation for some scanning
strategies. Another simplification concerned the foreground spectral
behaviour.  It was assumed that spectral parameters, such as the
synchrotron spectral index and dust emission properties, were
spatially-invariant. In the analysis of real data, however, one cannot
simply ignore these effects, and it is very important to investigate
their influence on the component separation results.

For real observations, each point in the sky is observed a different
number of times depending on the scanning strategy of the
instrument. In the case of simple scanning strategies, such as
constant latitude scans, the spin axis stays close to the plane of the
ecliptic. In this situation, pixels near the ecliptic pole are
observed several times more often than those near the ecliptic
plane. This uneven coverage leads to marked differences in the noise
rms per pixel across the sky. We also note that instrinsic gain fluctuations
in the instrument can also contribute to the variable noise rms.

Component separation using incomplete maps containing cuts (e.g. along
the Galactic plane) can be considered as an extreme case of varying
noise rms. One approach is to assume that the noise rms for 
pixels in the cut is formally infinite (or, equivalently, that they
have zero statistical weight). We show below that this method does
indeed allow the separation method to cope straightforwardly with
cuts. Moreover, this technique can in principle be applied to analyse
arbitrarily-shaped regions on the celestial sphere.

The assumed isotropy of the foreground spectral parameters over the
sky is another extreme simplification. For example, the synchrotron
spectral index varies in a wide range. Giardino \et (2002) calculated
the synchrotron temperature spectral index using three low-frequency
radio survey maps and found it to vary in the range 2.5$\le
\beta_{408/1420} \le$3.2.  Variation of the dust colour temperature
$T_{\rm dust}$ is more important for the {\sc Planck} experiment
because the HFI channels are quite sensitive to thermal dust
emission. Schlegel \et (1998) found $T_{\rm dust}$ to vary in the
range 16K to 20K, about a mean value $\langle T_{\rm dust}
\rangle$=18K assuming a single--component model.

The effect of a spatially-varying Galactic dust emissivity index
$\beta$ on the MEM reconstruction was investigated for case of a
flat-sky patch by Jones \et (2000).  Several dust sub-components with
different emissivities, but with the same colour temperature, were
included in the separation process, which provided a good
reconstruction of the components. More recently, Barreiro \et (2004)
used a combined real and harmonic space-based MEM technique to perform
a component separation on real data, in the presence of anisotropic
noise, cut-sky maps and spectral index uncertainties. Since this
approach requires multiple transitions between pixel and spectral
domains, however, the computation of the necessary spherical harmonic
transforms makes it is much slower than harmonic-space MEM, and hence
it was implemented only for low-resolution COBE data.

In this paper we will demonstrate how to extend the full-sky
harmonic-space MEM component separation method to take into account
anisotropic noise and variations in spectral parameters, by making use
of prior knowledge of the uneven sky coverage and the average value of
the spectral parameters.  The structure of the paper is as follows. In
a Section~\ref{noise} we summarise the basics of the MEM component
separation technique, describe the model of the microwave sky used in
the simulations, and review the impact of the non-isotropic noise on
the CMB reconstruction. In a Section~\ref{dust_temp} we will introduce
an approach for taking account of dust colour temperature variations
in the component separation process, describe the new microwave sky
model with variable $T_{\rm dust}$ and show the results for the
reconstruction tests made using different approximations. In
Section~\ref{discussion} we discuss the results and present our
conclusions.

\section{Anisotropic noise and incomplete sky coverage}
\label{noise}

Following the notation from previous articles (Hobson \et 1998; S02), 
a data vector of length $n_f$, which  contains the
observed temperature (or specific intensity) fluctuations at $n_f$
observing frequencies in any given direction $\hvect{x}$, can be defined as
\begin{equation}
d_\nu(\hvect{x}) = \int_{4\pi} B_\nu(\hvect{x}\cdot\hvect{x}') 
\sum_{p=1}^{n_c} F_{\nu p}\,s_p(\hvect{x}') \,{\rm d}\Omega'
+ \epsilon_\nu(\hvect{x})
\label{datadef}
\end{equation}
where $F_{\nu p}$ is the frequency response matrix, $B_\nu$ is the
beam profile for the $\nu$th frequency channel,
$\epsilon_\nu(\hvect{x})$ is the instrumental noise contribution in
the $\nu$th channel, and $s_p(\hvect{x})$ is the signal from $p$th
physical component. An integration is performed over the solid angle
$\Omega$.

For observations over the whole sky and for a Gaussian (or at least
circularly symmetric) beam profile at each frequency we can rewrite
the previous equation in matrix notation using the spherical harmonic
coefficients:
\begin{equation}
{\mathbfss d}_{\ell m} = {\mathbfss R}_\ell {\mathbfss a}_{\ell m}
+\bmath{\epsilon}_{\ell m},
\label{dataft2}
\end{equation}
where ${\mathbfss d}_{\ell m}$, ${\mathbfss a}_{\ell m}$ and
$\bmath{\epsilon}_{\ell m}$ are column vectors containing $n_f$, $n_c$
and $n_f$ complex components respectively. The response matrix
${\mathbfss R}_\ell$ has dimensions $n_f\times n_c$ and accommodates
the beam smearing and the frequency scaling of the components.

We assume that the anisotropic, uncorrelated pixel noise represents a
non-stationary random process in the signal domain with mean value
$\langle\epsilon_\nu(\hvect{x})\rangle=0$ and an rms that varies 
across the sky.  This leads to correlations between different
$(\ell,m)$ modes in the spherical harmonic domain.
Assuming that the instrumental noise is uncorrelated 
between different frequency channels, we have
\begin{equation}
\langle \epsilon_{\ell m}(\nu)\epsilon^\dagger_{\ell' m'}(\nu') \rangle
 = {\cal N}_{\ell m,\ell' m'}(\nu)\delta_{\nu\nu'}, \label{nmat}
\label{noise_cov}
\end{equation}
where the form of ${\cal N}_{\ell m,\ell' m'}(\nu)$ is discussed in
detail in Appendix A. In particular, for the case in which the
instrumental noise is uncorrelated between pixels (as we consider in
Section~\ref{model_rms}), we have
\begin{equation}
{\cal N}_{\ell m,\ell' m'}(\nu) 
\equiv  \Omega^2_{\rm pix} \sum_{p=1}^{N_{\rm pix}} Y_{\ell m}(\hvect{x}_p) 
Y^*_{\ell' m'}(\hvect{x}_p) \sigma_{\nu}^2(\hvect{x}_p),
\label{lmrms}
\end{equation}
where $\sigma_{\nu}^2(\hvect{x}_p) = \langle \epsilon_{\nu}^2
(\hvect{x}_p) \rangle$ is the noise variance in the $p$th pixel, 
$\Omega_{\rm pix}$ is the pixel area and $N_{\rm pix}$ is the total number of
pixels. For each frequency $\nu$, if we define the double indices
$i\equiv \ell m$ and $j\equiv \ell' m'$, we may consider the
quantities (\ref{lmrms}) as the components of the noise
covariance matrix $\bmath{\cal N}(\nu)$ in the spherical harmonic
domain. It is also shown in Appendix A that the elements of the
inverse of this matrix are given by
\begin{equation}
\left[\bmath{{\cal N}}^{-1}(\nu)\right]_{\ell m,\ell' m'} 
= \sum_{p=1}^{N_{\rm pix}} Y_{\ell m}(\hvect{x}_p) 
Y^*_{\ell' m'}(\hvect{x}_p) \frac{1}{\sigma_{\nu}^2(\hvect{x}_p)},
\label{lmrmsinv}
\end{equation}
where, in general, the notation $[\cdots]_{ij}$ denotes the $ij$th
element of the corresponding matrix.

Ideally, one would like to take into account the full noise covariance
matrix in performing the component separation. As discussed in
Stolyarov et al. (2002), however, it is not computationally feasible
to determine the entire vector, $\hat{\mathbfss a}$, containing the best
estimate of the harmonic coefficients of the physical components,
using all the elements of the data vector $\vect{\mathbfss d}$
simultaneously. Instead, a `mode-by-mode' approach is used, in which
any a priori coupling between different $(\ell,m)$ modes is neglected.
Taking the modes to be independent corresponds to assuming that the
likelihood and prior probability distributions factorise, such that
\begin{eqnarray}
\Pr({\mathbfss d}|{\mathbfss a}) & = & \prod_{\ell, m} \Pr({\mathbfss
d}_{\ell m}|{\mathbfss a}_{\ell m}), \label{likeprod}\\ 
\Pr({\mathbfss a}) & = &
\prod_{\ell, m} \Pr({\mathbfss a}_{\ell m}). \label{priorprod}
\end{eqnarray}
This offers an enormous computational advantage, since one can
maximise the posterior probability
\begin{equation}
\Pr({\mathbfss a}_{\ell m}|{\mathbfss d}_{\ell m}) \propto
\Pr({\mathbfss d}_{\ell m}|{\mathbfss a}_{\ell m}) \Pr({\mathbfss
a}_{\ell m})
\end{equation}
at each mode separately, 
where $\Pr({\mathbfss d}_{\ell m}|{\mathbfss a}_{\ell m})$ is the
likelihood and $\Pr({\mathbfss a}_{\ell m})$ is an entropic
prior. 

The factorisation (\ref{likeprod}), together with the assumption that
the instrumental noise is Gaussian,  implies that the likelihood
function is given by
\begin{equation}
\Pr({\mathbfss d}_{\ell m}|{\mathbfss a}_{\ell m}) \propto \exp
\left[-\chi^2({\mathbfss a}_{\ell m})\right],
\label{like1}
\end{equation}
where $\chi^2$ is the standard misfit statistic
\begin{equation}
\chi^2({\mathbfss a}_{\ell m}) = ({\mathbfss d}_{\ell m}-{\mathbfss
R}_\ell {\mathbfss a}_{\ell m})^\dag {\mathbfss N}^{-1}_{\ell m}
({\mathbfss d}_{\ell m}-{\mathbfss R}_\ell {\mathbfss a}_{\ell m}).
\label{like2}
\end{equation}
In this expression ${\mathbfss N}^{-1}_{\ell m}$ is the $n_f \times
n_f$ inverse noise covariance matrix for the $(\ell, m)$
mode, which can be different for each mode.
For the case in which the instrumental noise is uncorrelated between
frequency channels, it is a diagonal matrix and we take the $\nu$th
diagonal entry to be
\begin{equation}
\left[{\mathbfss N}^{-1}_{\ell m}\right]_{\nu\nu} =
\left[\bmath{{\cal N}}^{-1}(\nu)\right]_{\ell m,\ell m},
\label{invndef}
\end{equation}
where the right-hand side is obtained by setting $\ell'=\ell$ and
$m'=m$ in (\ref{lmrmsinv}).
One should note that we are considering {\em inverse} noise covariance
matrices in the expression (\ref{invndef}). This is clearly {\em not}
equivalent simply to setting $[{\mathbfss N}_{\ell m}]_{\nu\nu}
=[\bmath{{\cal N}}(\nu)]_{\ell m,\ell m}$ and 
then inverting the resulting matrix
(which reduces in this case to taking reciprocals of the diagonal
elements). 

The reason for using the definition (\ref{invndef}) is that it allows
for the straightforward analysis of cut-sky data.  We can consider the
missing areas in cut-sky maps as an extreme case of anisotropic noise
in which the noise rms of the pixels in the cut is formally infinite.
As can be seen from (\ref{lmrms}), this leads to elements of $\bmath{\cal
N}(\nu)$ which are also formally infinite, and hence this causes
problems for the analysis. One might try to avoid such difficulties
by, for example, setting the noise rms in the cut to be some arbitrary
large value.  This can itself cause problems of computational
accuracy, however, and one also has no a priori means of deciding to
which large value the noise rms in the cut should be set. Moreover,
the resulting component separation may depend on the value
adopted. Using (\ref{lmrmsinv}) instead, we see that formally infinite noise
rms in the cut is easily accommodated. It corresponds simply to
omitting the pixels in the cut from the summation in
(\ref{lmrmsinv}). Adopting this approach also means that there is no need to
apply any smoothing to the edges of the cut, or to perform the
analysis on some cut-sky harmonic basis. One is, in effect, still
performing the component separation over the whole sky, but not
constraining the solution in the cut region in any way. After the
component separation has been performed one may then apply whichever
cut is desired to the resulting component maps.

\subsection{Simulated {\sc Planck} observations}
\label{model_rms}

For our component separation tests, we prepared simulated {\sc Planck}
observations for 9 frequency bands at 30, 44, 70 GHz (LFI) and 100,
143, 217, 353, 545 and 857 GHz (HFI) with a resolution of 3.4 $arcmin$
(HEALPix parameter $N_{side}$=1024). The key parameters of the
satellite, such as beam FWHMs and average noise levels, were taken
from the relevant web-page\footnote{{\scriptsize
http://astro.esa.int/Planck/science/performance/perf\_top.html}}.  In
our simulations and separation tests, we assumed symmetrical Gaussian
beams.

\begin{figure}
\epsfig{file=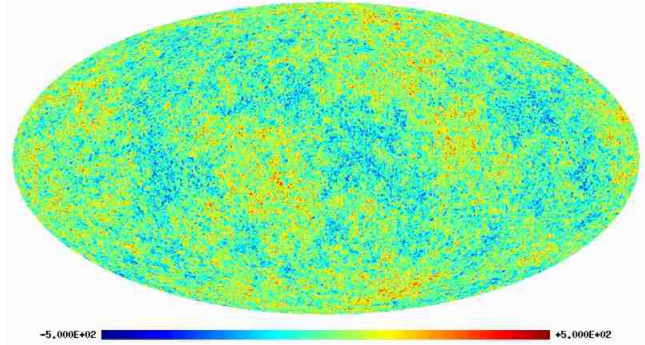,width=8.5cm}
\caption{The template of the primordial CMB anisotropies 
used in the simulations. The map is plotted in units of $\mu K$}
\label{cmb}
\end{figure}

The models of the astrophysical components used in the simulations are
the same as those presented in S02. We assume the presence of six
physical components of emission: primordial CMB, thermal and kinetic
Sunyaev-Zel'dovich (SZ) effects, and Galactic synchrotron, free-free
and thermal dust emission.  Fig.~\ref{cmb} shows the template of the
primordial CMB anisotropies used in the simulations.

\begin{figure}
\epsfig{file=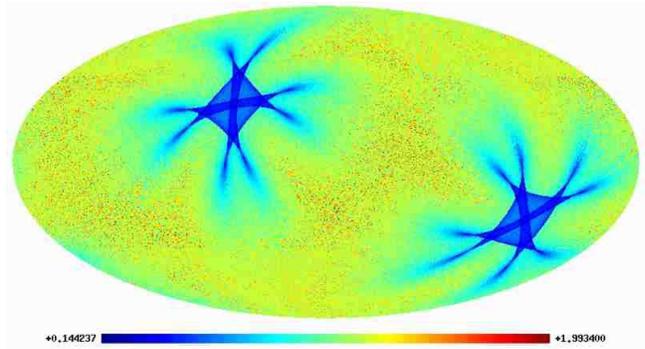,width=8.5cm}
\caption{The noise rms variation across the sky, in Galactic
coordinates, due to uneven coverage resulting from a 
sinusoidal-precession scanning strategy with $\theta = 90\degr$,
$\Delta \theta = 7\degr$, opening angle $\phi = 85\degr$ and
$n_{cycle} = 4$. The plot is scaled such that unity corresponds to
the average noise rms. Note that the noise rms decreases near the
ecliptic poles.\label{skycoverage}}
\end{figure}

Anisotropic, uncorrelated instrumental noise is simulated for the
sinusoidal-precession scanning strategy with $\theta = 90\degr$,
$\Delta \theta = 7\degr$, opening angle $\phi = 85\degr$ and
$n_{\rm cycle} = 4$.  The ecliptic latitude $\beta$ depends on the ecliptic
longitude $\lambda$ as $\beta=\Delta\theta \sin(n_{\rm cycle}\lambda)$.
A map of the corresponding sky coverage for this
scanning strategy is shown in Fig.~\ref{skycoverage}.  All frequency
maps are modelled with the same sky coverage for simplicity.

\begin{figure}
\epsfig{file=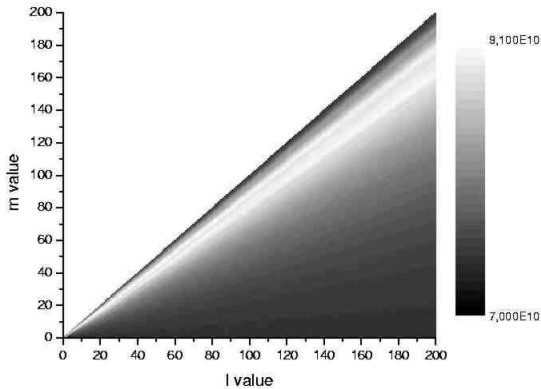,width=6.5cm,angle=-90}
\caption{The statistical weight $w_{\ell m}$ of each $(\ell,m)$ mode
for the 100 GHz HFI frequency channel assuming the scanning strategy
illustrated in Fig.~\ref{skycoverage}. In the case of uniform noise
$w_{\ell m} = 6.1\times 10^{10}$~(MJy/Sr)$^{-2}$ for this channel.}
\label{noise_alm}
\end{figure}

On calculating the `statistical weight' $w_{\ell m} \equiv
[\bmath{{\cal N}}^{-1}]_{\ell m,\ell m}$ for each $(\ell,m)$ mode using
(\ref{lmrmsinv}), we might expect some trends and variations along
both $\ell$ and $m$. For this particular scanning strategy, however,
with full sky coverage the variations are not very large. In
Fig.~\ref{noise_alm} we plot the relative amplitude of the statistical
weights for the 100 GHz channel on large scales ($\ell \leq 200$).  It
is clear from the plot that the deviations from the average level do
not exceed the 10 per cent level.

\begin{figure}
\epsfig{file=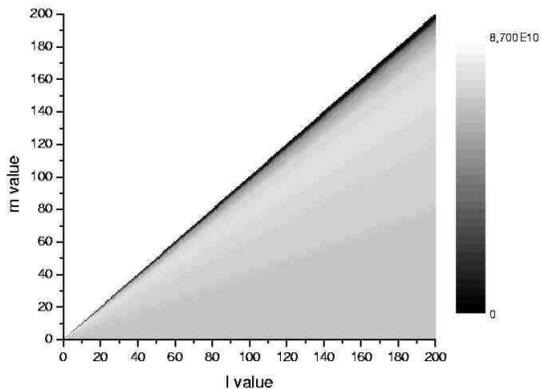,width=6.5cm,angle=-90}
\caption{As in Fig.~\ref{noise_alm}, but for a map containing a
 symmetric, constant-latitude cut of $\pm 10\degr$ along the Galactic
 plane.}
\label{cut_noise_alm}
\end{figure}

As mentioned above, a cut-sky map can be considered as an extreme case
of anisotropic noise. To test the accuracy of the component separation
in the presence of a cut, we also prepared a set of channel maps with
a symmetric, constant-latitude cut of $\pm 10\degr$ applied along the
Galactic plane. In the remainder of the map, we preserved the
anisotropic noise pattern resulting from the uneven coverage produced
by our assumed scanning strategy, but in the cut region the noise rms
was considered as formally infinite.  As might be expected, an extreme
anisotropy of this sort leads to more profound variations in the
statistical weight $w_{\ell m}$ of the modes. In
Fig.~\ref{cut_noise_alm}, we plot $w_{\ell m}$ for the 100 GHz channel
in this case. It is easy to see from the plot that the deviations from
the average level are significant.

\subsection{Results for all-sky data}
\label{results_rms}

\begin{figure}
\epsfig{file=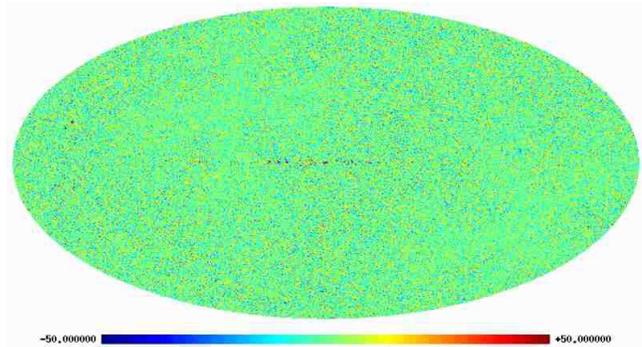,width=8.5cm}
\caption{The residuals of the CMB anisotropy reconstruction. 
The map is plotted in units of $\mu K$}
\label{cmbres}
\end{figure}

We first performed a component separation of simulated data with the
anisotropic noise distribution shown in Fig.~\ref{skycoverage},
assuming diagonal inverse noise covariance matrices ${\mathbfss
N}^{-1}_{\ell m}$ with elements given by (\ref{invndef}).  The
reconstructed map obtained for each physical component was of a
similar visual quality to those obtained in S02, and so they are not
presented here. Instead, we plot only the residuals of the
reconstruction of the primordial CMB component in Fig.~\ref{cmbres}.
We see that these residuals are essentially featureless, except for a
small band in the Galactic plane. We note, however, that the residuals
do contain a faint imprint of the noise rms distribution plotted in
Fig.~\ref{skycoverage}.

\begin{figure}
\epsfig{file=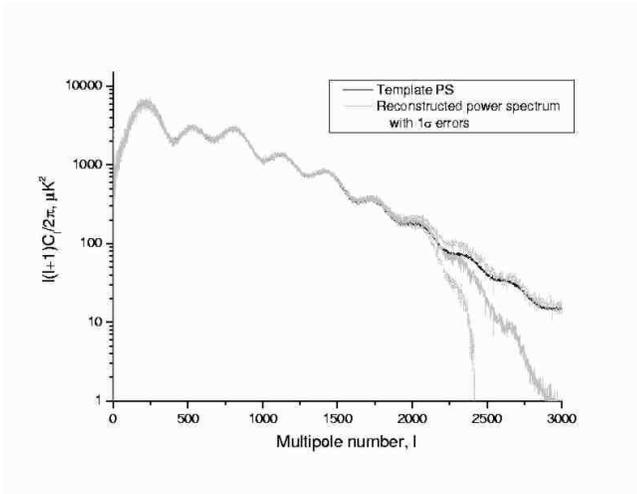,angle=-90,width=8.5cm}
\caption{Initial and unbiased reconstructed CMB power spectra for the
modelled data with non-uniform pixel noise. The 1$\sigma$ errors
on the reconstructed power spectrum are also shown.}
\label{CMB_ps1}
\end{figure}

In Fig.~\ref{CMB_ps1} we present the unbiased estimator (see S02) of
the CMB power spectrum, obtained from the reconstructed modes for this
component, and compare it with the power spectrum of the input
primordial CMB realisation used in the simulations. We see from the
figure that, even in the presence of realistic anisotropic noise, the
reconstructed CMB power spectrum follows the input spectrum out to
$\ell \approx 2500$, recovering the correct positions and heights of
the first seven acoustic peaks.
 
\begin{figure}
\epsfig{file=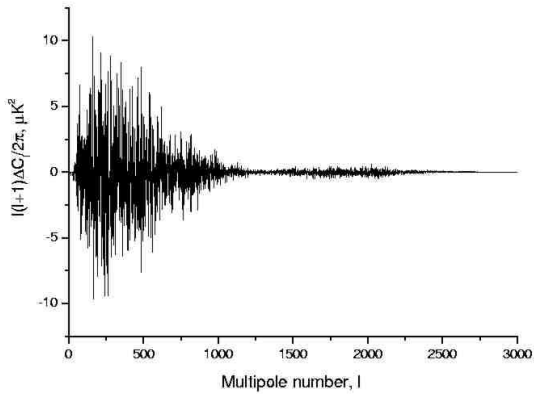,angle=-90,width=8.5cm}
\caption{The difference in the unbiased estimators
of the CMB power spectrum obtained assuming uniform noise
across the sky, and by taking into account noise rms variations.}
\label{CMB_DCl}
\end{figure}

In fact, even without making use of our prior knowledge of the noise
anisotropy, the effect of the spatially-varying noise rms on the
reconstructions is negligible.  In order to compare our new approach
with that presented in S02, we analysed the same simulated data once
more, but assuming that the noise was in fact isotropic.  The
resulting recovered maps were visually indistingushable from those
obtained using the approach outlined above. In Fig~\ref{CMB_DCl}, we
plot the difference in the unbiased estimators of the CMB power
spectrum obtained by the two methods. We see that the two approaches
yield almost identical results. This is not too surprising, since the
variation in the values of the statistical weight per $(\ell,m)$ mode
are quite small in the all-sky case, as illustrated in
Fig.~\ref{noise_alm}.

\subsection{Results for cut-sky data}

In the process of foreground removal from CMB data (as opposed to
component separation), it is common practice to impose some cut along
the Galactic plane prior to performing the analysis. In S02, it was
shown that such an approach is unnecessary, since accurate results
may be obtained by instead performing the component separation on
all-sky data, and then imposing a Galactic cut in the reconstructed
CMB map, if required, prior to any further analysis. Nevertheless,
the question of analysing cut-sky data sets might occur, for example,
in the event of instrument failure, and it is thus of interest to
investigate how to accommodate this complication.

\begin{figure}
\epsfig{file=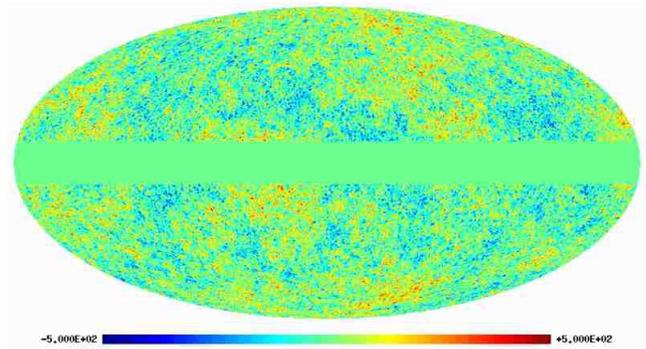,width=8.5cm}
\caption{The reconstruction of the CMB anisotropy in the cut-sky case.
The map is plotted in units of $\mu K$}
\label{cmb_cut}
\end{figure}

\begin{figure}
\epsfig{file=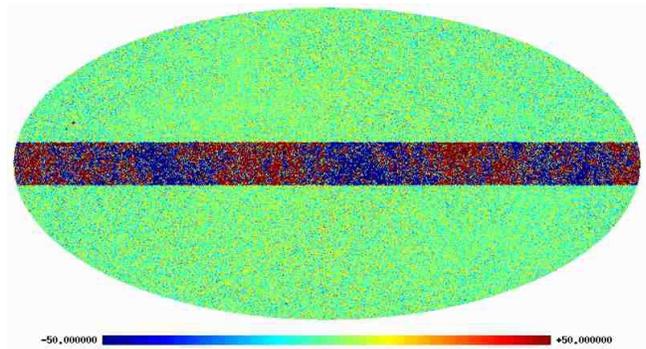,width=8.5cm}
\caption{The residuals of the reconstruction in cut-sky case.
The map is plotted in units of $\mu K$}
\label{cmb_cut_resid}
\end{figure}

To test our approach of modelling cut-sky data as an extreme case of
anisotropic noise, we performed a component separation using a set of
frequency maps with $\pm 10\degr$ symmetrical, constant-latitude cut
along the Galactic plane, within which the pixels were set to zero. 
In Fig.~\ref{cmb_cut} we show the reconstruction of the primordial CMB
component, and the corresponding residuals are shown in
Fig.~\ref{cmb_cut_resid}. 

The reconstructed physical components were not explicitly restricted
to be zero in the cut. Instead, as explained above, the reconstruction
was simply not constrained by the data in the cut region.  As a result
the reconstruction of spherical harmonic modes that lie predominantly
in the cut will be prior driven. Since the maximum of the entropic
prior occurs for zero signal, such modes will have zero amplitude in
the reconstruction. This does not mean, however, that the
reconstructed temperature map will be precisely zero in the cut.
Modes that are not restricted to the cut region will have amplitudes
that are constrained to be non-zero by data on the un-cut part of the
sky.  Such modes may, in general, contribute to the reconstructed
pixel values in the cut region.

We note from Fig.~\ref{cmb_cut} and Fig.~\ref{cmb_cut_resid} that, by
using our approach, the quality of the CMB reconstruction outside the
cut is unaffected by the presence of the cut. In particular, we see
that the reconstruction residuals are featureless over the un-cut part
of the sky, and show no identifiable structure even directly adjacent
to the cut region. In making this observation, it should also be
remembered that no smoothing was applied to the edges of the cut prior
to the analysis.

\begin{figure}
\epsfig{file=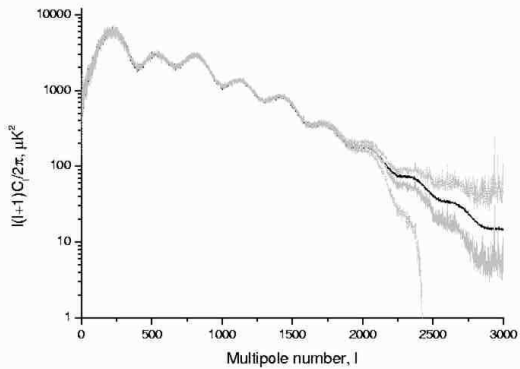,angle=-90,width=8.5cm}
\caption{Initial and unbiased reconstructed CMB power spectra for
data containing a Galactic cut. The 1$\sigma$ errors
on the reconstructed power spectrum are also shown.}
\label{cmb_cut_ps}
\end{figure}

Finally, in Fig.~\ref{cmb_cut_ps}, we plot the unbiased estimator (see
S02) of the CMB power spectrum, obtained from the reconstructed modes
for this component, and compare it with the power spectrum of the
input primordial CMB realisation used in the simulations. We see that,
even in the presence of the cut, the reconstructed CMB power spectrum
follows the input spectrum out to $\ell \approx 2500$.

\section{Spatially-varying foreground spectra}
\label{dust_temp}

Another major problem for harmonic-space component separation methods,
such as MEM and Wiener filtering, is the implicit assumption in
(\ref{datadef}) that the emission from each physical component can be
factorised into a spatial template at some reference frequency $\nu_0$
and a frequency dependence, so that
\begin{equation}
I_\nu(\hvect{x}) = \sum_{c=1}^{n_c} s_c(\hvect{x}) f_c(\nu).
\end{equation}
This clearly represents the idealised case in which the spectral
parameters of each component do not vary with direction on the sky.
In terms of the detailed operation of the harmonic-space MEM algorithm
(and Wiener filtering), this assumption is quite central. The MEM
approach seeks to find the most probable values for the harmonic
coefficients ${\mathbfss a}$ of the astrophysical components, given the
observed data and an entropic regularisation prior on the solution. At
each step in the optimisation, the predicted data corresponding to the
current best estimate of ${\mathbfss a}$ is compared (through the Gaussian
likelihood function) with the real data. In calculating the predicted
data at each frequency, one must determine the contribution of the
physical components at this frequency by scaling the solution from the
reference frequency $\nu_0$ using the conversion matrix $F_{\nu c}$.
This matrix determines the frequency behaviour of the components and
its coefficients are calculated in accordance to some model of
spectral scaling -- power law for the synchrotron, modified blackbody
emission law for the dust emission, and so on.  The scaling is taken to
be the same for all $(\ell,m)$ modes, which implies that the
parameters defining the spectral scaling do not vary across the sky.

The assumption of spatially-invariant spectral parameters is
reasonable for the primordial CMB and the two SZ effects. The thermal
SZ effect does depend on the electron temperature of the clusters
$T_e$, which can reach 10--15 keV; this dependence is rather weak, but
can be used to determine $T_e$ during component separation, as will be
explored in a forthcoming paper.  For the Galactic components,
however, our assumption is clearly not valid.  For example, as
discussed in Section~\ref{intro}, the spectral indices of the
synchrotron and thermal dust emission are known to vary considerably
across the sky.  In the case of {\sc Planck} observations, variation
in the synchrotron spectral index will not cause severe problems. Even
in the lowest frequency channel at $\nu=30$ GHz, the synchrotron
emission is quite weak. A similar conclusion may be drawn for
free-free emission.  Unfortunately, variations in the spectral
parameters of thermal dust emission can give rise to severe
difficulties in performing component separation for {\sc Planck} data.

In the previous simulations presented in S02, the thermal dust
emission was assumed to follow a simple single-component grey-body model
with spectral emissivity index $\beta=2$ and mean dust temperature
$\langle T_{\rm dust} \rangle=18$ K. It is easy to show that variations
in these parameters at 20---30 per cent level can
lead to differences of the same factor (in specific intensity units)
when scaling from the reference frequency $\nu_0=300$ GHz to the
highest HFI channel at $\nu=857$ GHz. This can cause huge errors
in the prediction of dust emission which can reach up to $10^5$ MJy/Sr
in some regions of the galactic plane. As a result, incorrectly
assuming the emissivity and dust temperature to be spatially-invariant
can severely reduce the accuracy with which all the physical
components are recovered. The dominant cause of these difficulties is
in fact the uncertainty in the dust temperature, which is known to
vary in the range 5--25 K. It is therefore necessary to take proper account
of the variation in the dust temperature to obtain reliable component
separation results.

Using the simplest single-component grey-body model as an example, the
scaling with frequency of the dust emission can be defined, in terms
of specific intensity, by
\begin{equation}
I_{\nu}(\hvect{x})=I_{\nu_0}(\hvect{x})\frac{\nu^\alpha B(\nu ,
T(\hvect{x}))} {\nu_0^\alpha B(\nu_0 , T(\hvect{x}))} =
I_{\nu_0}(\hvect{x}) F_{\nu c}(T(\hvect{x})),
\label{dust_scaling}
\end{equation}
where $I_{\nu}(\hvect{x})$ is the dust specific intensity at frequency
$\nu$ in the direction $\hvect{x}$, $T(\hvect{x})$ is dust colour
temperature in this direction, $\beta$ is the emissivity spectral
index, which is assumed to be uniform over the sky, and $B(\nu, T)$ is
the blackbody function. The quantity $F_{\nu c}(T(\hvect{x}))$ is the
appropriate element of the conversion matrix for the dust component,
which depends on the position on the sky.

Since (\ref{dust_scaling}) takes the form of a product of two
functions in the spatial domain, we cannot easily pass to the
spectral domain, because this product will turn into a convolution
over the whole range of the spherical harmonics. In order to work in
terms of harmonic coefficients, we instead begin by expanding 
(\ref{dust_scaling}) around the mean dust temperature $T_0$ to obtain
\begin{equation}
I_{\nu}(\hvect{x}) \approx I_{\nu_0}(\hvect{x})F_{\nu c}(T_0) + 
I_{\nu_0}(\hvect{x}) \Delta T(\hvect{x}) 
\left.\frac{\partial F_{\nu c}}{\partial T} \right\vert_{T=T_0} + \cdots
\label{dust_scaling_approx}
\end{equation}
where $\Delta T(\hvect{x})$ is the deviation from the mean temperature
in a given direction. We have not included further terms in the
expansion for the sake of brevity, but these may be written
down trivially. Given our Taylor series expansion, 
it is now possible to use our standard
formalism to reconstruct several separate (but highly correlated)
fields -- intensity at the reference frequency $I_{\nu_0}$ and
intensity-weighted temperature fields $I_{\nu_0} \Delta T$, $I_{\nu_0}
(\Delta T)^2$, and so on, to obtain an increasingly accurate
approximation to $I_\nu(\hvect{x})$.

A similar approach may also be used for accommodating spatial
variation in any other foreground spectral parameter, such as the dust
emissivity or synchrotron spectral index. One may also treat the
electron temperature of SZ clusters in the same way.  It should be
noted, however, that the number of terms used in the Taylor expansion
cannot simply increase without limit. Typically, one is constrained
such that the total number of fields to be reconstructed does not
exceed the number of frequencies at which observations are made.

\subsection{Simulated {\sc Planck} observations}
\label{model_temp}

The simulated {\sc Planck} observations considered here are similar to
those described in Section~\ref{model_rms}. We assume six physical
components and simulate nine frequency maps, smoothed with appropriate
FWHMs and add anisotropic pixel noise. The only difference here is
that the thermal dust emission has a spatially-varying temperature
distribution.

\begin{figure}
\epsfig{file=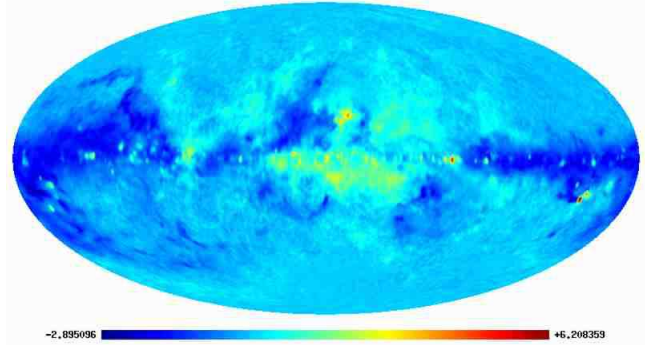,width=8.5cm}
\caption{The map of the temperature variations across the sky in K
(Schlegel \et 1998). The original map from Schlegel \et 1998, with
$\langle T_{\rm dust} \rangle=18K $, was scaled to $\langle T_{\rm
dust} \rangle = 16.2K$ and the mean value has been subtracted.}
\label{dust_template}
\end{figure}

In simulating the thermal dust emission, it is more accurate to use a
two-component best-fit thermal dust model, with the dominant component
having a mean temperature of $\langle T_1 \rangle = 16.2$ K and a small
contribution of cold dust with $\langle T_2 \rangle =9.4$ K
(Finkbeiner \et 1999). The two components are fully correlated and
their temperature is related by $T_2 = 0.352 \,T_1^{1.18}$, so we have
only one effective temperature parameter. A template for the
temperature $T_1(\hvect{x})$ across the sky was constructed from the
colour temperature of Schlegel \et (1998).  The latter was constructed
assuming a single-component dust model with the uniform emissivity
index of $\beta=2$ and has the mean temperature $\langle T_{\rm dust}
\rangle=18$K. This map was simply scaled to $\langle T_{\rm dust}
\rangle = 16.2K$ to obtain the $T_1(\hvect{x})$ map for the
two-component model used in our simulations (see
Fig.~\ref{dust_template}).

For a multi-component dust model, the dust emission, in specific
intensity units, scales with frequency as
\begin{eqnarray}
I_{\nu}(\hvect{x}) & = & I_{\nu_0}(\hvect{x})
\frac{\sum_k{f_k B(\nu , T_k(\hvect{x}))q_k(\frac{\nu}{\nu_0})^{\alpha_k}}}
{\sum_k {f_k B(\nu_0 , T_k(\hvect{x}) ) q_k }} \nonumber \\
& = & I_{\nu_0}(\hvect{x}) F_{\nu c}(T(\hvect{x})),
\label{dust_scaling_multi}
\end{eqnarray}
where $k=1,2,...$ labels the dust component, $f_k$ is a normalisation
factor for the $k$th component and $\sum_k{f_k}=1$. The terms $q_k
(\nu / \nu_0)^{\alpha_k}$ represent the relative emission efficiency for
each component (see Finkbeiner et al (1999) for details).  For the
two-component best-fit model used in our simulations, $f_1$=0.0363,
$\alpha_1$=1.67, $\alpha_2$=2.70 and $q_1/q_2$=13.0 . We can expand
the multi-component dust model around the mean temperature an
analogous manner to that given in (\ref{dust_scaling_approx}) for the
single component case.

\begin{figure}
\epsfig{file=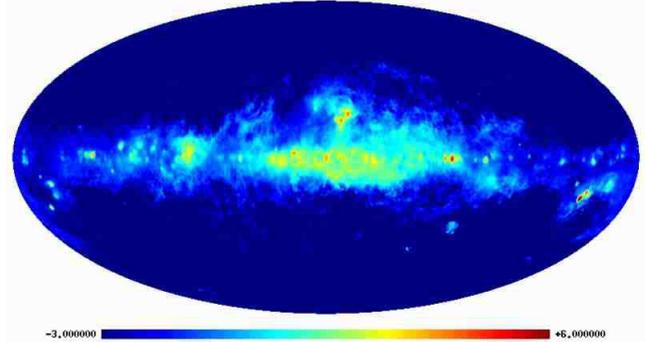,width=8.5cm}
\caption{The reconstruction of the colour dust temperature variations
$\Delta T_{\rm dust}$. The map is plotted in units of $K$}
\label{dust_recon}
\end{figure}

\subsection{Component separation results}
\label{results_temp}

Several separation tests were performed to investigate the approach
outlined above.  In our first test, we recreated the method used in
S02, in which the dust temperature was assumed to be
spatially-invariant (the zeroth-order approximation), and so only
intensity fields $I^{(k)}_{\nu_0}$ $(k=1,2,\ldots,6)$ for the six
physical components were reconstructed at the reference frequency
$\nu_0$.  In the second test, we attempted to reconstruct 7
components, which included $I^{(k)}_{\nu_0}$ for $k=1,2,\ldots,6$, but
also the intensity-weighted dust colour temperature field $I_{\nu_0}
\Delta T$ (the first order of the expansion).  Finally, in a third
test, 8 components were reconstructed, namely $I^{(k)}_{\nu_0}$ for
$k=1,2,\ldots,6$, together with $I_{\nu_0} \Delta T$ and $I_{\nu_0}
(\Delta T)^2$, thereby accounting for dust temperature variations up
to second order.

\subsubsection{Quality of the CMB reconstruction}

\begin{figure*}
\begin{center}
\centerline{\epsfig{file=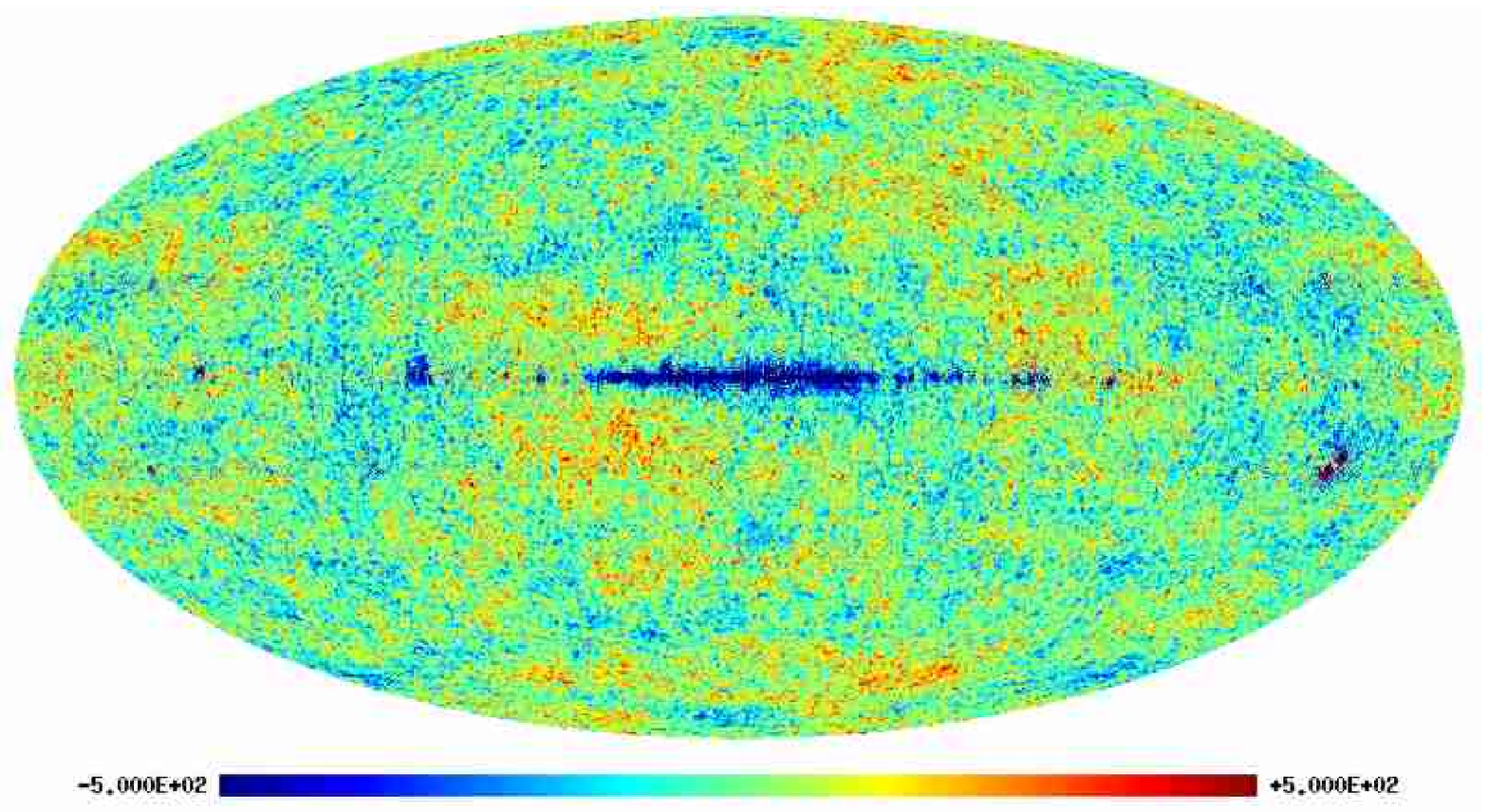,width=8.5cm}
\quad 
\epsfig{file=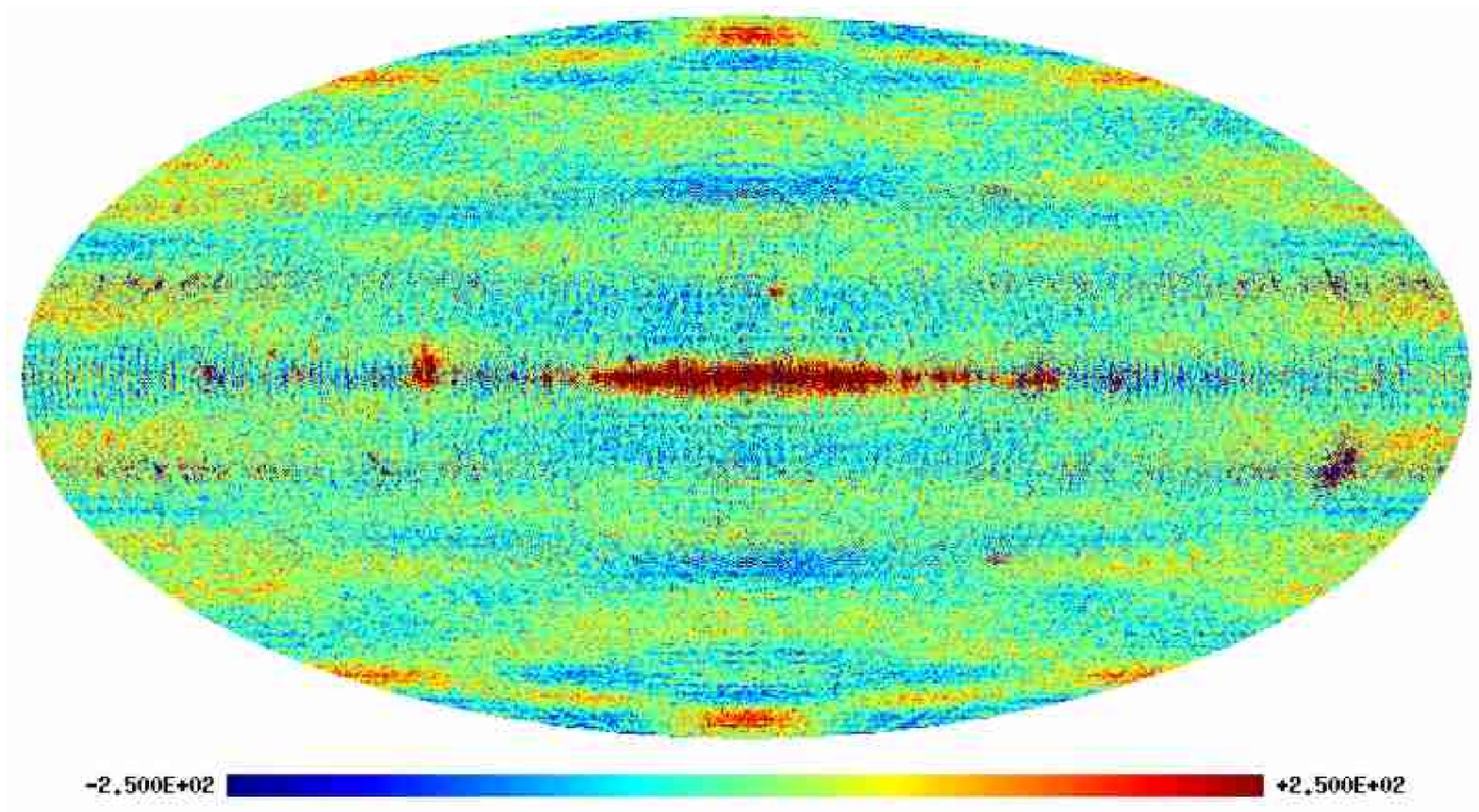,width=8.5cm}}
\centerline{
\epsfig{file=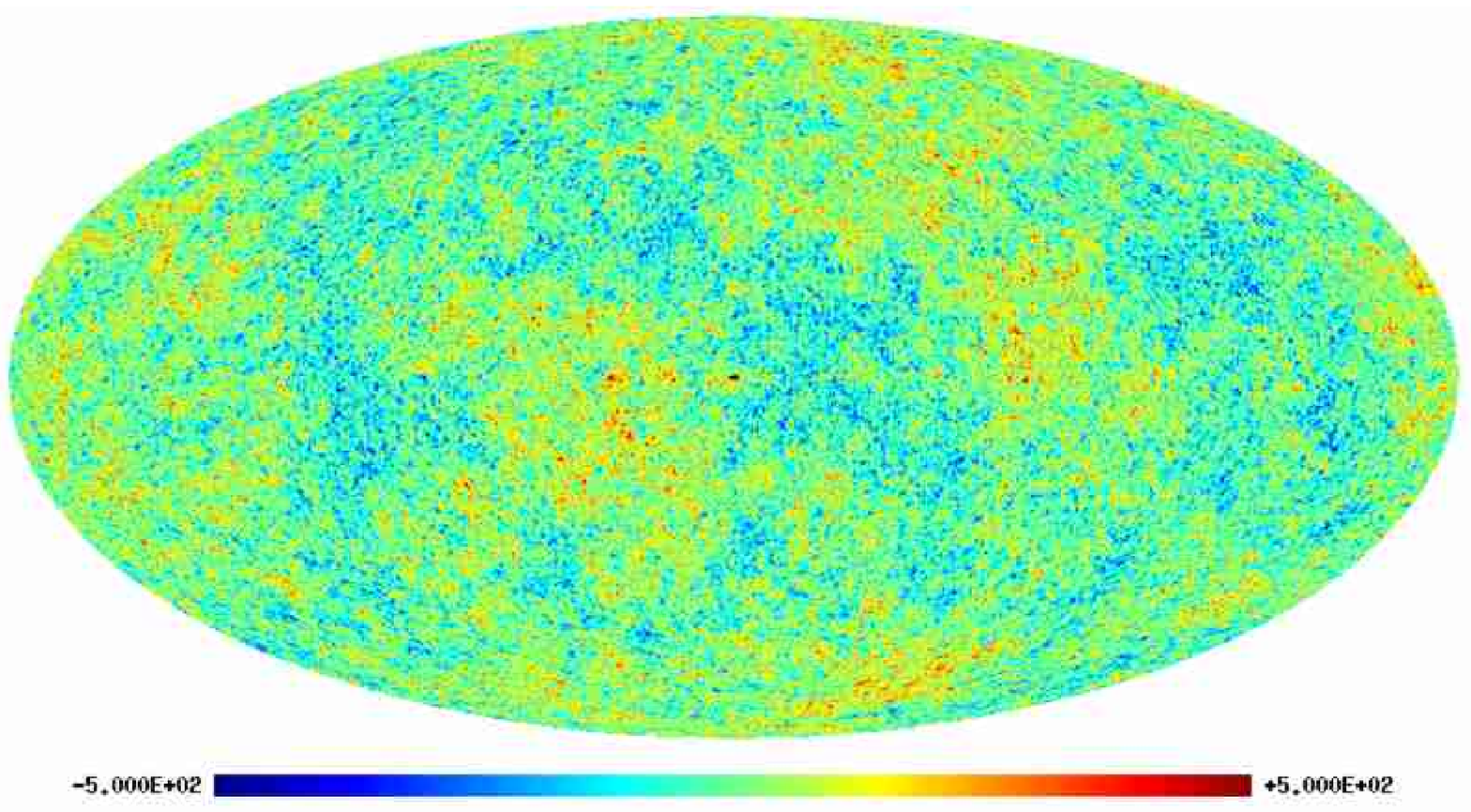,width=8.5cm}
\quad 
\epsfig{file=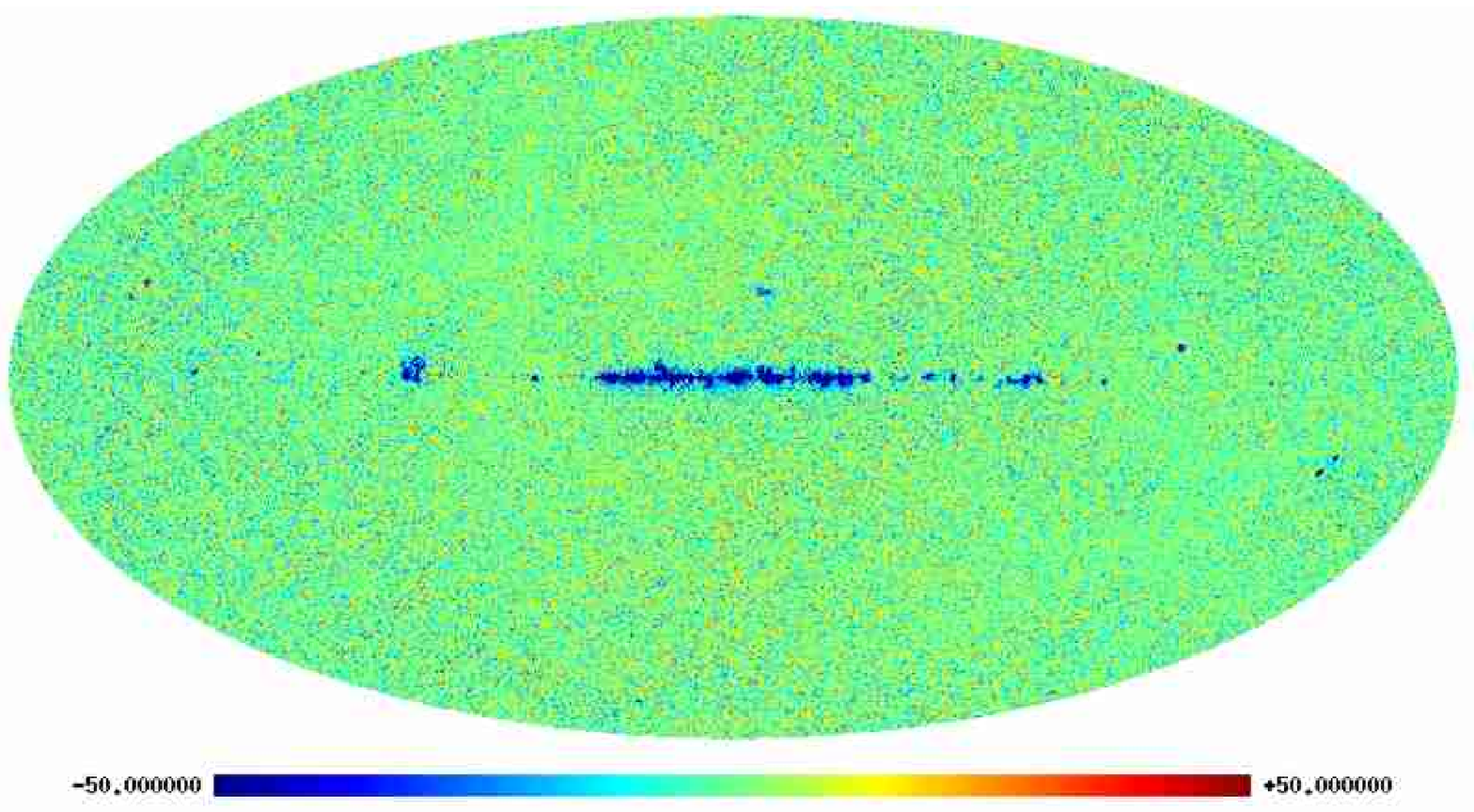,width=8.5cm}}
\centerline{
\epsfig{file=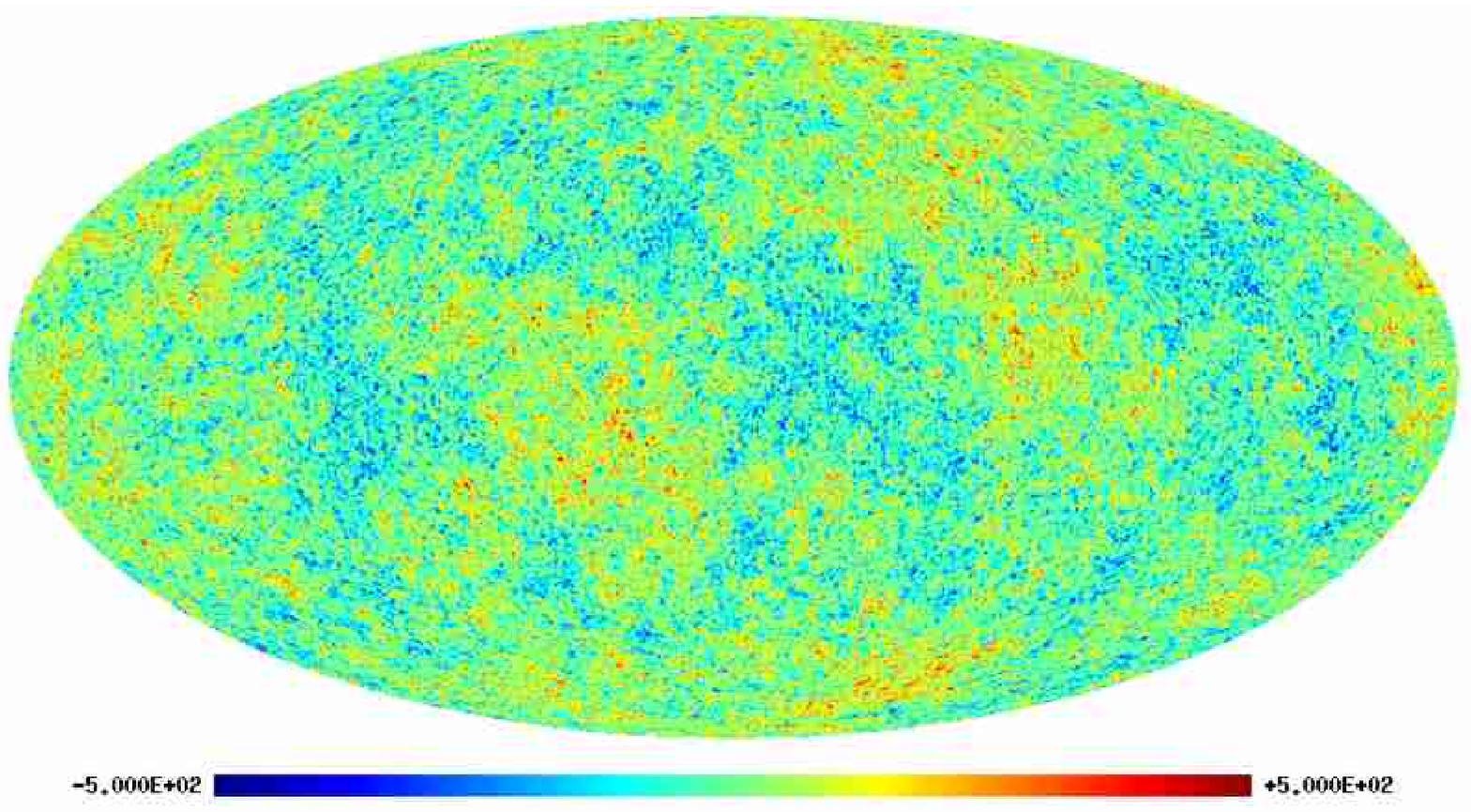,width=8.5cm}
\quad 
\epsfig{file=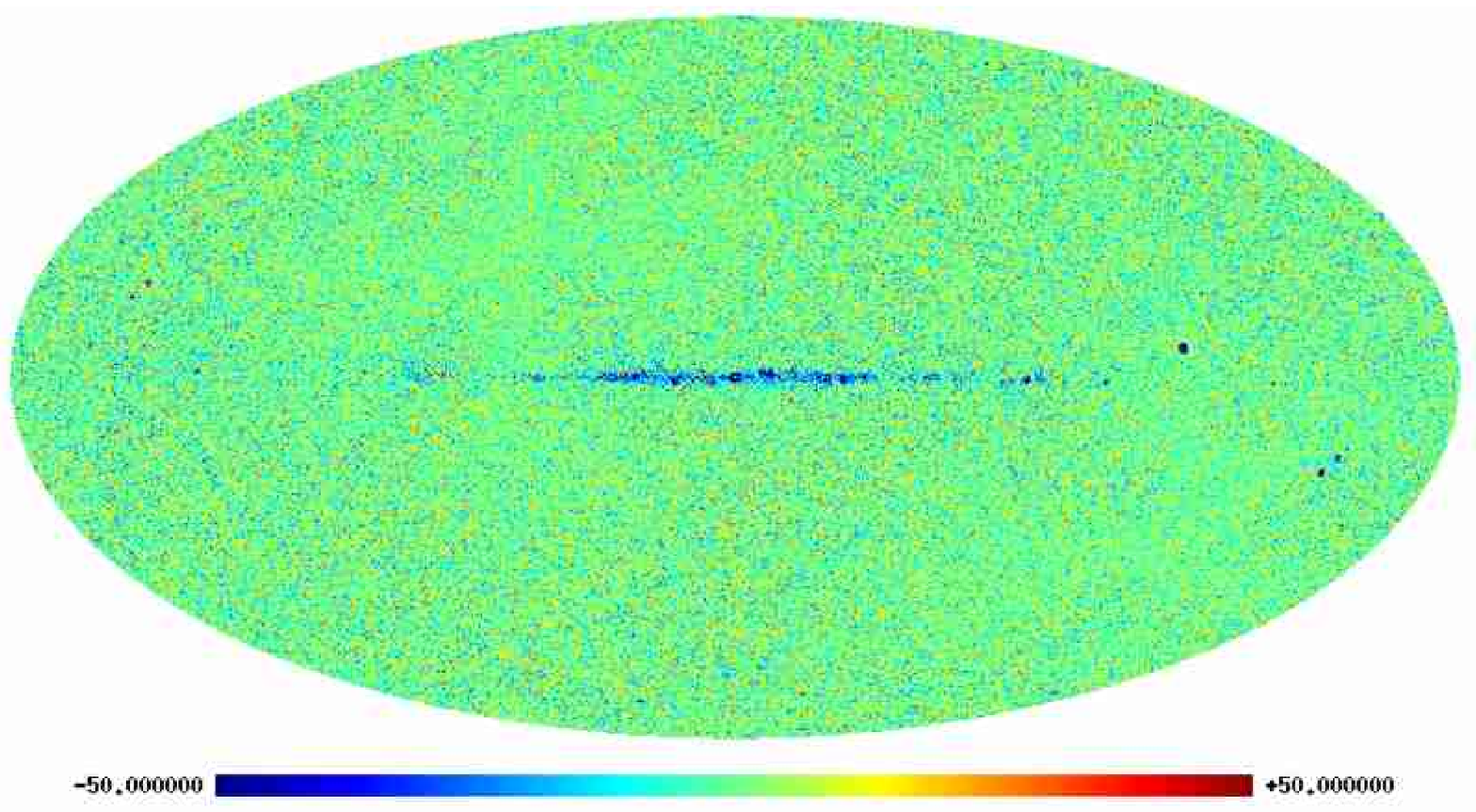,width=8.5cm}}
\caption{Reconstructions of the primordial CMB component (left-hand
column) and the corresponding residuals (right-hand column) for the
three separation tests: without any accounting for $T_{\rm dust}$
variations across the sky (top panel); accounting for variations up to
first order (middle panel); and accounting for variations up to second
order (bottom panel).  All maps are plotted in units of $\mu$K.}
\label{cmb_recon}
\end{center}
\end{figure*}

The reconstruction of the primordial CMB component and the
corresponding residuals are shown in Fig.~\ref{cmb_recon} for all
three separation tests. It is clear from the figure that the
separation assuming $T_{\rm dust}$ to be constant over the sky has a
very strong residual signal along the plane of Galaxy on the CMB map,
where the dust component is strong. Moreover, we see that the
reconstruction outside the Galactic plane is also badly contaminated;
this is discussed further in Section~\ref{discussion}.  We thus
conclude that the practice adopted in previous component separation
analyses of assuming a spatially-invariant dust temperature leads to
CMB reconstructions of unacceptably poor quality. Although we do not
plot them here, the reconstructions of the other physical components
are found to contain similar artefacts both inside and outside the
Galactic plane.
\begin{figure*}
\centerline{
\epsfig{file=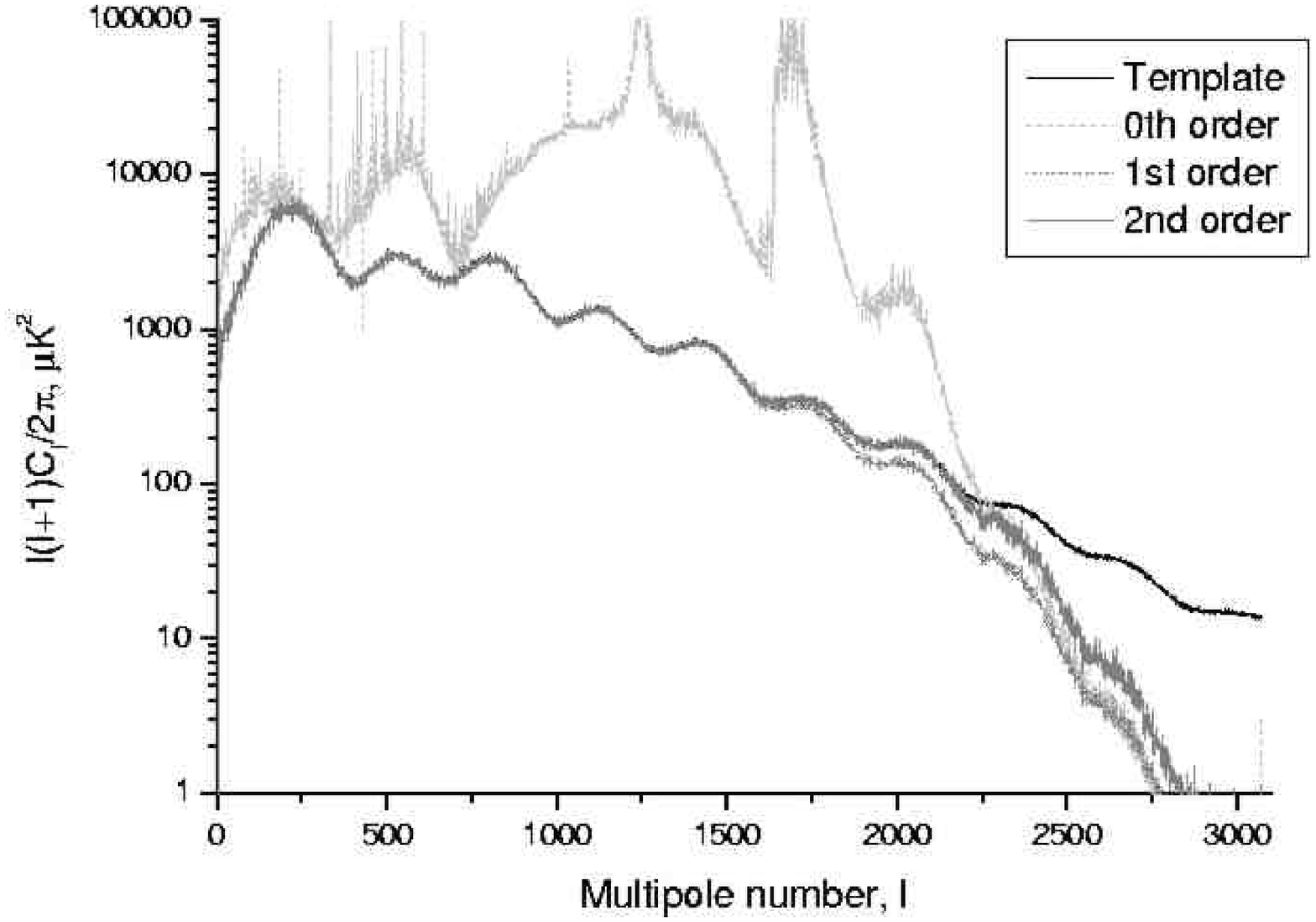, width=6.5cm,angle=-90}
\qquad
\epsfig{file=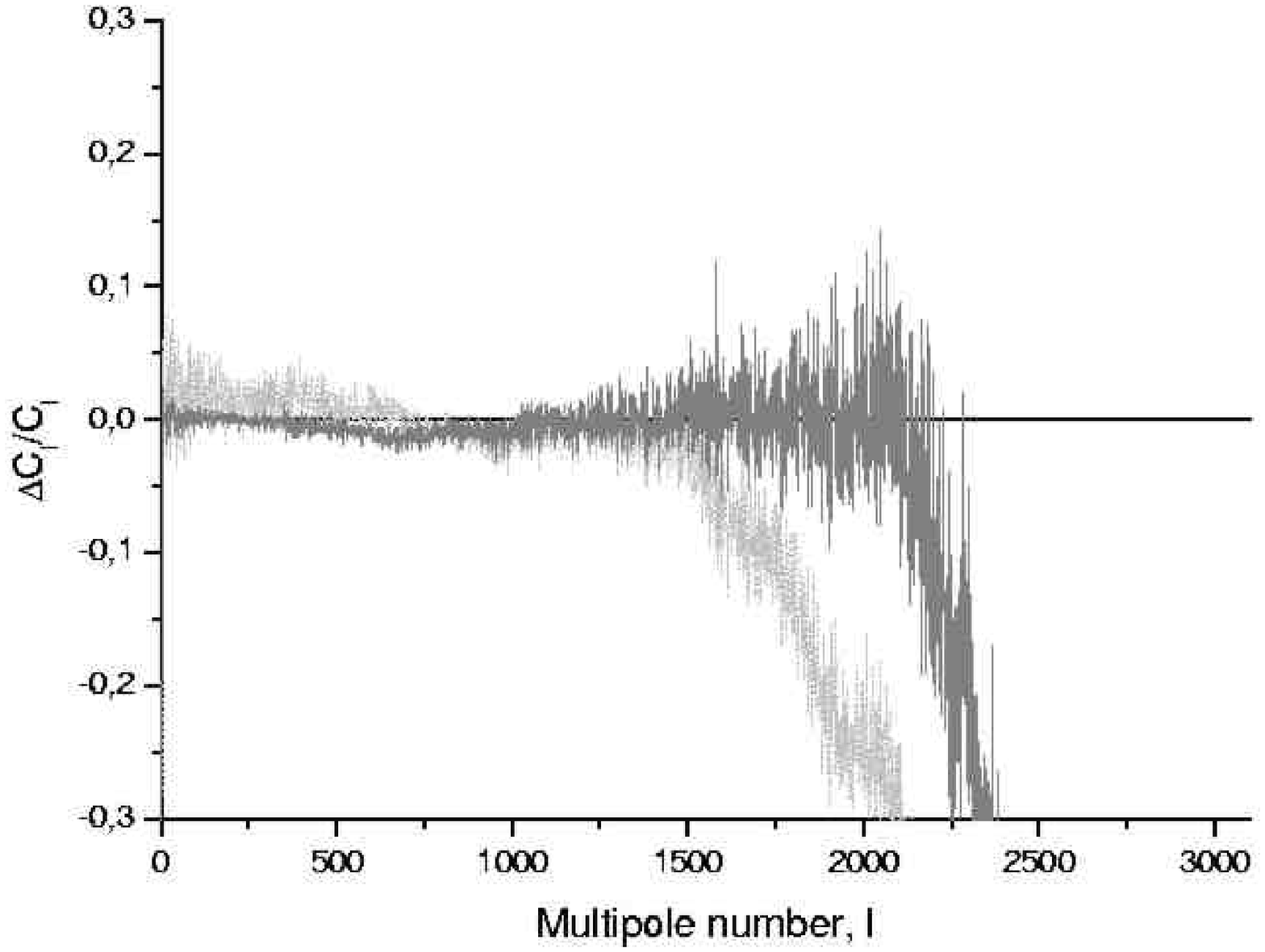, width=6.5cm,angle=-90}}
\caption{(Left) The input and unbiased reconstructed CMB power spectra
  for the three separation tests. (Right) The relative difference
  between the input and reconstructed power spectra $(C^{\rm rec}_l -
  C^{\rm template}_l)/C^{\rm template}_l$ for the second and third
  cases.}
\label{cmb_ps_tdust}
\end{figure*}

Fortunately, the reconstruction errors are significantly reduced even
when the dust temperature variations are accounted for only up to
first order in the expansion (\ref{dust_scaling_approx}). We see that
outside Galactic regions of high dust emission, the CMB residuals in
this case are essentially featureless. Nevertheless, the Galactic
plane still contains significant dust contamination. The quality of
the CMB reconstruction can be improved still further by accounting for
$T_{\rm dust}$ variations up to second order. In this case, the
remaining contamination of the CMB residuals in the Galactic plane is
significantly reduced, to a level comparable to that obtained in S02 in
the analysis of simulated data in which $T_{\rm dust}$ was taken not
to vary.

A more quantitative comparison of the quality of the CMB
reconstructions is given in Fig.~\ref{cmb_ps_tdust} (left panel), in
which the unbiassed estimates (see S02) of the angular power spectra
of the three reconstructions are plotted in relation to the input
power spectrum. In each case, the power spectrum has been calculated
using the full sky reconstruction.  As anticipated from
Fig.~\ref{cmb_recon}, the recovered power spectrum in the zeroth-order
approximation is very poor due to the significant dust
contamination. The recovered power spectrum in the first-order
approximation is significantly improved. It follows the input spectrum
out to $\ell \approx 1700$, recovering the positions and heights of
the first 5 acoustic peaks. Beyond this point, the presence of the 6th
and 7th acoustic peaks is inferred, but the recovered spectrum begins
to underestimate the true power level. In the second-order
approximation, the power spectrum recovery is improved still further,
and the positions and heights of the first 7 acoustic peaks are
accurately recovered before the true power level is again
underestimated. The differences between the first and second-order
recovered power spectra are further illustrated in the right panel of
Fig.~\ref{cmb_ps_tdust} in which the ratio $(C^{\rm rec}_l - C^{\rm
template}_l)/C^{\rm template}_l$ is plotted for these two cases. The
improved accuracy and extended range of validity of the second-order
approximation is easily seen.

\subsubsection{Reconstruction of the dust temperature distribution}

For the second and third separation tests, the actual dust
colour temperature variations can be calculated as
\begin{equation}
T_{\rm dust}(\hvect{x}) = \langle T_{\rm dust} \rangle + (I_{\nu_0} \Delta
T)^{\rm dust}_{\hvect{x}}/ I^{\rm dust}_{\nu_0}(\hvect{x}),
\label{dust_temp1}
\end{equation}  
where $I^{\rm dust}_{\nu_0}$ is the reconstruction of dust intensity
at the reference frequency $\nu_0$ assuming the average colour
temperature $\langle T_{\rm dust} \rangle$, and $(I_{\nu_0} \Delta
T)^{\rm dust}$ is the reconstruction of intensity-weighted temperature
variation field. The result is shown in
Fig.~\ref{dust_recon}. Comparing with Fig.~\ref{dust_template}, we see
that the reconstuction has faithfully recovered the main features of
the input map. It is clear, however, that $\Delta T_{\rm dust}$
variations are better restored around the Galactic plane. This is not
surprising, since we have actually reconstructed the
intensity-weighted $T_{\rm dust}$ field, which can be recoverd more
accurately in areas with higher dust intensity, which are obviously
located predominantly in the Galactic plane.

\section{Discussion and conclusions}
\label{discussion}

In this paper, we have demonstrated an approach that allows one to
account for the spatial variations of the noise properties and
spectral characteristics of foregrounds in the harmonic-space
maximum-entropy component separation technique.

In Section~\ref{results_rms}, we show that the impact of a
realistic level of the anisotropic noise on the quality of the
foreground separation is quite small, at least for the simple scanning
strategy assumed in generating our simulated observations. This can be
explained by the fact that the rms noise level differs from the
average level only in relatively small areas around ecliptic
poles. This leads to variations in the noise level at each harmonic
mode that differ from the average value by only a few per cent.
We also illustrate in Section~\ref{results_rms} that the it is
possible to perform harmonic-space component separation on cut-sky
maps by treating the cut as an extreme example of anisotropic noise.
This appraoch has the advantage of not requiring one to smooth the
edges of the cut with some apodising function prior to the analysis.

In Section~\ref{dust_temp}, we show that the variation of spectral
parameters of foregrounds may be taken into account by a method of
succesive approximations based on a series expansion of the
corresponding intensity field around the mean value of the parameter.
In particular, we investigate the effect of dust colour temperature
variations on the quality of the component separation, focussing on
the reconstruction of the CMB.  We show that realistic dust
temperature variations lead to severe contaminaton of the CMB
reconstruction if, in the separation process, the dust temperature is
assumed not to vary. This contamination is concentrated in the
Galactic plane, but significant artefacts exist at high Galactic
latitudes. The poor quality of the reconstruction outside the Galactic
plane is a result of performing the reconstruction mode-by mode in
harmonic space. The inaccurate model of the dust emission leads to
errors in the determined amplitudes of a wide range of spherical
harmonics in the CMB reconstruction. Many of these modes do not lie
predominatly in the Galactic plane region, but contribute to the
reconstruction over the whole sky. 

\begin{figure*}
\begin{center}
\centerline{\epsfig{file=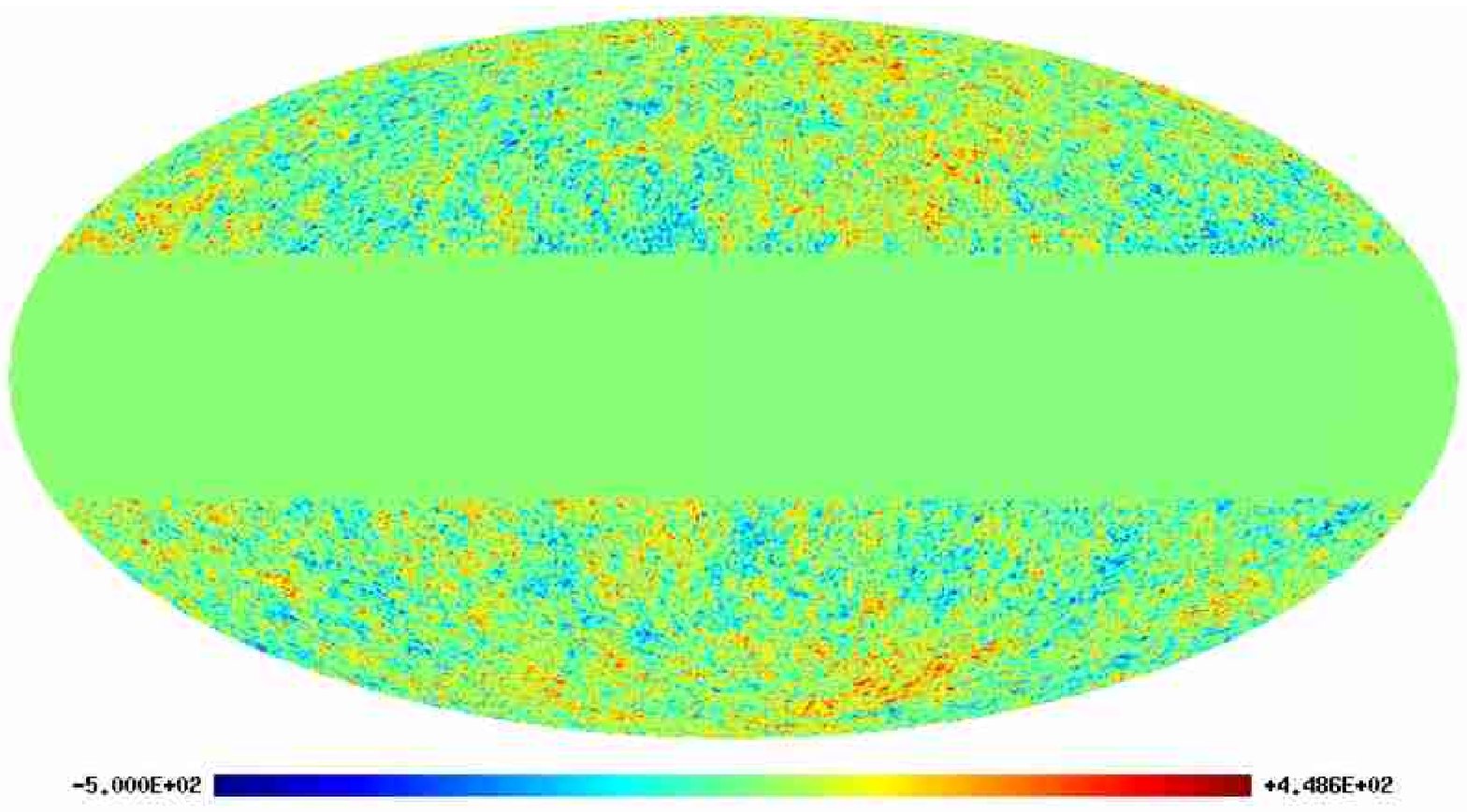,width=8.5cm}
\quad 
\epsfig{file=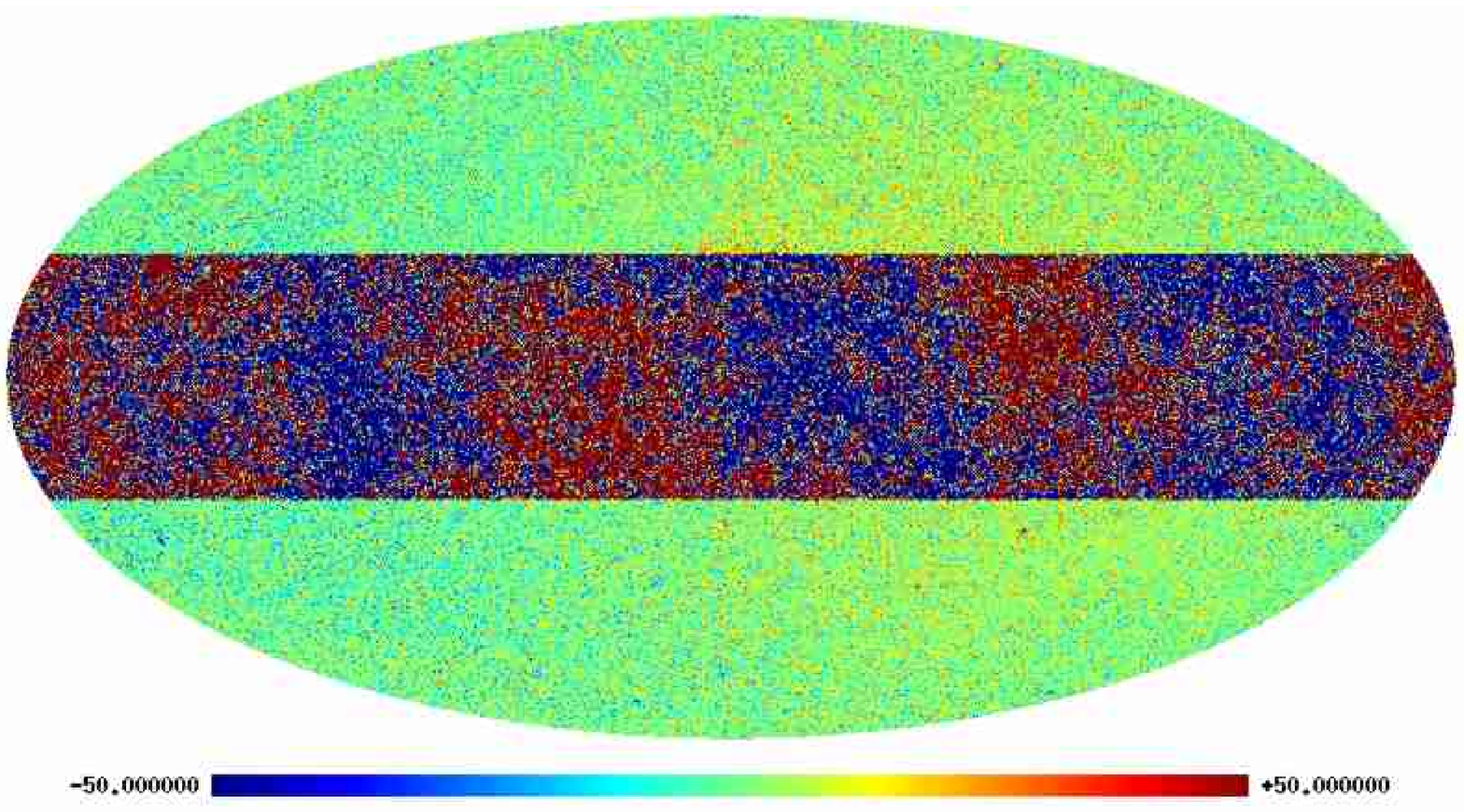,width=8.5cm}}
\caption{The reconstruction (left) and residuals (right) of 
the CMB for the case
  plotted in Fig.~\ref{cmb_recon} (upper row), but with a Galactic cut of 
$\pm 25\degr$. The maps are plotted in units of $\mu K$.}
\label{cmb_dust_cut}
\end{center}
\end{figure*}

If one is content simply with removing foregrounds from the CMB,
rather than performing a component separation, then one could apply a
Galactic cut prior to the analysis.  In Fig.~\ref{cmb_dust_cut},
we show the CMB reconstruction and residuals respectively obtained by
applying a Galactic cut of $\pm 25\degr$ to each simulated frequency
map used in Section~\ref{model_temp}, and assuming a constant dust
temperature across the sky. 
The noise rms in the cut region was
assumed to be formally infinite in the manner discussed in
Section~\ref{noise}.  We see from the figures that the quality of the
reconstruction is significantly improved as compared with the case in
which no Galactic cut was applied, which was shown in
Fig.~\ref{cmb_recon} (top row). In particular, we note that the
residuals now contain no obvious artefacts outside of the cut region.
Thus, even assuming an inaccurate dust model, one can still recover a
reasonable reconstruction of the CMB outside of the Galactic plane.

It is clearly not possible, however, to perform an acceptable all-sky
component separation for the {\sc Planck} experiment by assuming
constant dust spectral parameters, even using all 9 frequency
channels. Nevertheless, taking account of dust temperature variations
up to first order in the series expansion significantly improves the
CMB reconstruction to an acceptable level. This reconstruction quality
is still further improved by including second-order terms and is then
comparable to that obtained for the ideal case presented in Stolyarov
\et (2002), where the simulated observations assumed no dust
temperature variation across the sky. Moreover, an accurate
reconstruction of the dust temperature variation is obtained over the
whole sky.

Finally we note that the approach for dealing with spatially-varying
spectral parameters described in this paper can also be applied to
other foregrounds to yield, for example, maps of the synchrotron
spectral index. The method can also be used to reconstruct the
electron temperature in clusters from their thermal SZ effect, as will
be discussed in a forthcoming paper.

\section{Acknowledgements}
Some of the results in this paper have been derived using HEALPix
(G\'orski, Hivon and Wandelt 1999) package. The authors thank Mark
Ashdown for many useful conversations regarding component separation
and Daniel Mortlock for supplying anisotropic noise models.  Numerical
calculations were performed on the National Cosmology Altix~3700
Supercomputer funded by HEFCE and PPARC and in cooperation with
Silicon Graphics.  RBB thanks the Ministerio de Ciencia y
Tecnolog\'{\i}a and the Universidad de Cantabria for a Ram\'on y Cajal
contract.

\appendix

\section{The noise covariance matrix in the spherical harmonic domain}
\label{app1}

We consider a single frequency map for which $n_p$ is the 
instrumental noise in the $p$th pixel. We denote the spherical harmonic
coefficients of the noise by $\epsilon_{\ell m}$. 

The elements of the noise covariance matrix in the harmonic domain are
given by
\begin{eqnarray}
{\cal N}_{\ell m,\ell' m'} 
& = & \langle \epsilon_{\ell m} \epsilon^\ast_{\ell' m'} \rangle \nonumber \\ 
& = & \langle
\Omega_{\rm pix} \sum_p \bigl (Y_{\ell m}(p) n_p \bigr ) ~ 
\Omega_{\rm pix} \sum_{p'}
\bigl ( Y^\ast_{\ell' m'}(p') n_{p'} \bigr ) \rangle \nonumber \\ 
& = & \Omega^2_{\rm pix} \sum_p \sum_{p'} Y_{\ell m}(p) Y^\ast_{\ell' m'}(p') 
\langle n_p n_{p'} \rangle \nonumber \\ 
& = & \Omega^2_{\rm pix} \sum_p 
Y_{\ell m}(p) Y^\ast_{\ell'  m'}(p) \sigma^2_p
\label{rms_per_mode}
\end{eqnarray}
where $Y_{\ell m}(p)$ denotes the value of the corresponding
spherical harmonic at the $p$th pixel centre, 
$\Omega_{\rm pix} = 4\pi / N_{\rm pix}$ and, in the last line,
we have assumed that the noise is uncorrelated between pixels, so
that $\langle n_p n_{p'} \rangle = \sigma^2_p \delta_{pp'}$. 
We note that in the special case in which the noise is statistically 
isotropic, $\sigma^2_p = \sigma^2$ for all pixels, and we obtain
\begin{eqnarray}
{\cal N}_{\ell m,\ell' m'} 
& = & \Omega_{\rm pix} \sigma^2
\Bigl [\Omega_{\rm pix} \sum_p Y_{\ell m}(p)
Y^\ast_{\ell' m'}(p) \Bigr ] \nonumber \\
& = & \Omega_{\rm pix} \sigma^2 \delta_{\ell\ell'}\delta_{mm'}.
\label{white_noise_sigma}
\end{eqnarray}

It is convenient to define the double indices $i \equiv \ell m$ and $j
\equiv \ell' m'$, and regard (\ref{rms_per_mode}) as the elements
of the noise covariance matrix $\bmath{\cal N}$. One may then write
\begin{equation}
\bmath{\cal N} = \Omega_{\rm pix}^2 
{\mathbfss Y}^\dagger {\mathbfss N} {\mathbfss Y},
\end{equation}
where ${\mathbfss N}$ is the noise covariance matrix in the pixel
domain, with elements $N_{pp'} = \langle n_p n_{p'} \rangle$, and we have 
defined the transformation matrix ${\mathbfss Y}$ with
elements $Y_{pi} = Y_i(p)$.

Using the fact that transformation matrix has the useful property
\begin{equation}
{\mathbfss Y}^\dagger {\mathbfss Y} = \frac{1}{\Omega_{\rm pix}}
\mathbfss{I} = {\mathbfss Y}{\mathbfss Y}^\dagger,
\end{equation}
it is straightforward to show that 
$\bmath{{\cal N}}\bmath{{\cal N}}^{-1} 
= {\mathbfss I} = \bmath{{\cal N}}^{-1}\bmath{{\cal N}}$, where the
inverse noise covariance matrix in the spherical harmonic domain is
given by
\begin{equation}
\bmath{{\cal N}}^{-1} = {\mathbfss Y}^\dagger
{\mathbfss N}^{-1} {\mathbfss Y}.
\end{equation}
In the special case that the instrumental noise is uncorrelated
between pixels, ${\mathbfss N}^{-1}$ is diagonal and we obtain
\begin{equation}
\left[\bmath{{\cal N}}^{-1}\right]_{\ell m, \ell' m'}
= \sum_p Y_{\ell m}(p) Y^*_{\ell' m'}(p) \frac{1}{\sigma_p^2}.
\end{equation}

\bsp

\label{lastpage}

\end{document}